\documentclass[12pt]{iopart}
\usepackage{iopams}  
\begin{document}

\title[5th LISA symposium]{Gravitational Waves
from binary black holes in the extreme mass ratio regime:
Self-force calculations}

\author{Carlos O. Lousto}

\address{Department of Physics and Astronomy,
and Center for Gravitational Wave Astronomy,
The University of Texas at Brownsville, Brownsville, Texas 78520, USA}

\begin{abstract}
We present a report on the progress made in the development of
computational techniques to evaluate the gravitational radiation
generated by a particle orbiting a massive black hole to second
pertrubative order.
\end{abstract}




\section{Introduction}

It is currently believed that most of the galaxies
host supermassive black holes (SMBH) in their cores \cite{Merritt:2004gc}.
Even our own galaxy host a black hole of almost 4 million solar masses
\cite{Weinberg:2004nj} in its center. In this scenario, the most
likely event is that stars and other compact objects in the
neighborhood of the SMBH will inspiral-down and plunge toward this
central massive black hole.

 From the Theoretical point of view the two body problem in General
Relativity remains one of the more interesting problems still
unsolved. While comparable mass binary black holes are the most
efficient astrophysical system regarding emission of gravitational
radiation, they represent also the most challenging problem for
numerical relativity. Current estimates show that during the final
plunge of two nearly equal mass black holes the system radiates as
much energy as in the whole previous inspiral history of the binary
\cite{Baker00b,Baker:2001nu,Baker:2001sf,Baker:2002qf,Baker:2004wv}.

For the Laser Interferometric Space Antenna (LISA), the small mass
ratio regime is particularly interesting since numerous sources will
be 'visible' with huge signal to noise ratios \cite{Gair:2004iv}. This
scenario is tractable using perturbation theory of black holes, but
proved to be a more challenging problem than initially expected, until
relatively recently, when decisive progress toward the complete
solution to the problem has been made, some of which we will briefly
report here.

\section{The problem and its solution}

The theory to compute radiation from an orbiting particle around a
black hole has been developed in the 70's
~\cite{Regge57,Zerilli70,Teukolsky73}. Soon after it was realized that
correcting the geodesic of the particle by its radiation formally lead
to infinities. In fact, if we try a straightforward solution to the
case of a particle represented by a Dirac delta in orbit around a
black hole we can write the field equations of general relativity as
\begin{equation}
G_{\mu\nu}[h_{\alpha\beta}]=T_{\mu\nu}=
m \int u_\mu u_\nu \delta[x^\alpha-x_p^\alpha(\tau)]/\sqrt{-g}\,d\tau
\end{equation}
where $h_{\alpha\beta}$ represents the perturbations around the larger
black hole background metric, $g^{(0)}_{\mu\nu}$ (with determinant
$g$).  $x_p^\alpha(\tau)$ is the (parametrized by the proper time of the
particle, $\tau$) coordinate
location of the particle (with mass $m$) orbiting the large hole and
$u^\mu={dx^\mu}/{d\tau}$ is its four-velocity.

Once we solve for the perturbed metric we can correct the trajectory
of the particle by requesting it to be a geodesic of the background
{\it plus} the perturbed metric
\begin{equation}
\frac{d^2x^\mu}{d\tau^2}+\Gamma^\mu_{\alpha\beta}\,\frac{dx^\alpha}{d\tau}
\,\frac{dx^\beta}{d\tau}=0.
\end{equation}
The problem arises when trying to evaluate the connection term
$\Gamma^\mu_{\alpha\beta}$ on the trajectory of the particle since it
can be readily shown that the metric behaves as
\begin{equation}
h_{\mu\nu}\sim \frac{m\,H_{\mu\nu}}
{\sqrt{\left(g_{\alpha\beta}+u_{\alpha}u_{\beta}\right)
\left(x^\alpha-x_p^\alpha(\tau)\right)
\left(x^\beta-x^\beta_p(\tau)\right)}}
\end{equation}
i.e., diverging as the inverse of the distance to the particle.
So, the small perturbations assumption breaks down near the particle.

In order for us to extract any physically meaningful result this
problem needs to be regularized. This proved to be a very challenging
problem for over 25 years until 1997 when Mino, Sasaki and
Tanaka~\cite{Mino:1997nk} published their seminal work on the
computation of the self-force a particle traveling in a curved
spacetime exerts on itself. They computed this by using two
alternative methods. First using matched asymptotic expansions of the
perturbed background of the larger black hole by the presence of the
particle and the spacetime around the small particle tidally distorted
by the larger hole. By matching this two expansions in a buffer zone
of radius $d$ around the particle, with $m<<d<<M$, one finds that
requiring the consistency of the two expansions leads to the equations
of motion. The second derivation is done via a Brehme-DeWitt approach
by deriving a 'conserved' rank-two symmetric tensor and integrating
its divergence over the interior of the world tube surrounding the
particle to obtain the equations of motion.

The self-force (defined as the correction to the background geodesic)
takes then the form
\begin{equation}
F^\mu_{full}(x)=m\nabla^{\mu\beta\gamma}h_{\beta\gamma}=
F^\mu_{direct}+F^\mu_{tail}
\end{equation}
where $F^\mu_{direct}$ is computed from contributions that propagate
{\it along} the past light cone and $F^\mu_{tail}$ has the
contributions from {\it inside} the light cone, product of the
scattering of perturbations due to the motion of the particle in
the curved spacetime created by the larger black hole.

The self-force is then computed by taking the limit
$F^\mu_{self}=F^\mu_{tail}(x\to x_p)$ where
\begin{equation}\label{MiSaTaQuWa}
F^\mu_{tail}=mu^\alpha u^\beta \int_{-\infty}^{\tau_-}
\left\{\frac12\nabla^\mu G^-_{\alpha\beta\gamma'\delta'}-
\nabla_\beta {G^-{_\alpha}}^{\mu}_{\ \gamma'\delta'}-
\frac12 u^\mu u^\lambda \nabla_\lambda G^-_{\alpha\beta\gamma'\delta'}
\right\} u^{\gamma'} u^{\delta'} d\tau'
\end{equation}
and where $G^-_{\alpha\beta\gamma'\delta'}$ is the retarded Green's
function. Primed indices denote evaluation at the particle's location
while unprimed denotes evalutation at the nearby field point $x$
$(u^\alpha u^\beta$ are obtained by parallel propagation.)  Finally,
$\tau_-$ represents the proper time of a point on the worldline that
intersects the particle's past lightcone of the field point $x$.
As a result, one finds that the divergent terms mentioned above do not
affect the motion of the particle.

Independently, Quinn and Wald~\cite{Quinn:1997am} come up with the
same equation of motion (\ref{MiSaTaQuWa}) (usually referred to as
'MiSaTaQuWa' force), but using an axiomatic approach:\par
\noindent
i) comparison axiom: Identifying the four-velocity $u^\mu$ and the
four-acceleration at points $P_1$ in a manifold
$(M_1,g^{(1)}_{\mu\nu})$ and $P_2$ in a manifold
$(M_2,g^{(2)}_{\mu\nu})$ via Riemann normal coordinates one find that
the difference between forces is given by
\begin{equation}
f_{(1)}^\mu-f_{(2)}^\mu=\lim_{r\to0}\left\{
\left(\frac12\nabla_{(1)}^\mu h^{(1)}_{\alpha\beta}-\nabla_{(1)\beta}\,h^{(1)\ \mu}_\alpha\right)
-\left(\frac12\nabla_{(2)}^\mu h^{(2)}_{\alpha\beta}-\nabla_{(2)\beta}\,h^{(2)\ \mu}_\alpha\right)
\right\}
\end{equation}
ii) Flat spacetime axiom: If $(M,g_{\mu\nu})$ is the Minkowski
spacetime the perturbed metric is the half advanced and half retarded
solution and the force vanishes
\begin{equation}
h_{\alpha\beta}=\frac12\left[h^+_{\alpha\beta}+h^-_{\alpha\beta}
\right],\ {\rm and}\ f^\mu=m\,u^\alpha\nabla^{(0)}_\alpha\,u^\mu=0.
\end{equation}

It is worth mentioning here a newer approach
\cite{Detweiler:2002mi} to the problem which decomposes the full
fields into a 'regular' (which is different from the 'tail' part) and
a 'singular' part. It can be proved that only the regular part
contributes to the self-force, producing the same 'MiSaTaQuWa' result,
and that the regular (metric) fields satisfy the vacuum equations of
General Relativity (with no source terms due to the presence of the
particle).

The self-force calculation thus provides a local correction to the
trajectory of the particle taking into account radiative and
nonradiative effects. For a global method that takes into account the
radiative effects only see, for instance Ref.~\cite{Hughes:2001jr},
that makes use of the energy and momentum balance of the radiation
emitted to infinity and onto the black hole horizon to
correct the particle trajectory over orbital cycles.

\section{The problems of the solution}

The proof that the equation of motion 
\begin{equation}
m\left(\frac{d^2x^\mu}{d\tau^2}+\Gamma^{(0)\,\mu}_{\alpha\beta}\,\frac{dx^\alpha}{d\tau}\,\frac{dx^\beta}{d\tau}\right)=F^\mu_{self}
\end{equation}
leads to the correct, finite description of the particle due to its own
field represent an enormous leap forward in our comprehension of the
effects of gravitational radiation. Yet, expression
(\ref{MiSaTaQuWa}) for the self-force is formal and it was not
immediately ready for implementation. It took a couple of years to
develop the formalism that allowed explicit computations. The key
observation, arrived to simultaneously by several authors
\cite{Barack:1999wf,Lousto99b,Burko:1999zy}, was
to tackle the nonrotating, i.e.  Schwarzschild black hole, background
case first to make use of its spherical symmetry to expand the force
into spherical harmonics.  For each multipole $(\ell,\tt{m})$, the full
force is finite, and the divergent nature of the problem only reveals
upon summing over $(\ell,\tt{m})$.

In the mode sum regularization method the force can be decomposed as
(wehere we performed alrady the sum over the modes $\tt{m}$)
\begin{equation}
F_{self}^\mu=\lim_{x\to x_p}\sum_{\ell}\left[F_{full}^{\mu\,\ell}(x)-F_{direct}^{\mu\,\ell}(x)\right]
\end{equation}
which upon addition and subtraction of  $A^\mu\,L+B^\mu+C^\mu/L$
(which leads to divergences upon summing over $\ell$)
\begin{eqnarray}
F_{self}^\mu&=&\sum_{\ell}
\left[F_{full}^{\mu\,\ell}(x_p)-(A^\mu\,L+B^\mu+C^\mu/L)\right]\nonumber\\
&-&\sum_{\ell}\left[F_{direct}^{\mu\,\ell}(x_p)-(A^\mu\,L+B^\mu+C^\mu/L)\right]
\end{eqnarray}
where for convenience we defined $L=\ell+1/2$. The terms in brackets
now lead to well defined, convergent series. The sum over the last
bracket is a remainder normally denoted as $D^\mu$.

In a notable piece of work, the regularization parameters $A^\mu,
B^\mu, C^\mu$ and $D^\mu$ have been computed {\it analytically}, first
for the Schwarzschild \cite{Barack:2001gx} and then for the Kerr
backgrounds \cite{Barack:2002mh}.

The missing part is now the full force, $F_{full}^{\mu\,\ell}(x_p)$,
that has to be computed by numerically integrating the (perturbed)
General Relativity field equations to obtain
$F_{full}^{\mu\,\ell}(x)$. For a nonrotating large black hole, we can
just solve the wave equations for the perturbations traveling on the
curved Schwarzschild background, i.e. the Zerilli and Regge-Wheeler
equation for the two polarizations of the radiation. We then know how
to reconstruct the metric in the Regge-Wheeler gauge
\cite{Lousto99b}.

We successfully followed this procedure in the case of a particle
starting from rest at a finite distance from the large black hole
\cite{BL02a}. In addition to the exact (numerical) computation of the
self-force, we obtained an analytic approximation
\begin{equation}\label{Fapprox}
F_{self}^r=F_{self}^{r\,\ell=0}+F_{self}^{r\,\ell=1}+
\sum_{\ell=2}^{\infty} \left\{-\frac{15}{16}m^2\frac{E^2}{r^2}
\left(E^2+\frac{4M}{r}-1\right)\frac{1}{L^2}+{\cal O}(L^{-4})\right\}
\end{equation}
where $M$ stands for the mass of the large hole and $(m,E)$ are the
mass and orbital energy of the particle respectively. Note also that
the force will have \cite{Lousto00a} only negative, even powers of $L$
what makes the sum quickly convergent and provides an excellent
approximation if we evaluate the first few lower multipoles
numerically.

After this successful computation we proceeded to compute the
astrophysically more appealing scenario of a particle in circular
orbit around a nonrotating black hole. In this case we encountered a
fundamental difficulty: The Regge-Wheeler gauge does not lead to $C^0$
metric coefficients at the particle location as for headon orbits.
Moreover, in general, a Dirac's delta behaviour appear in the metric
coefficients at $x_p$. Besides, the
regularization parameters have been computed in the harmonic gauge.

We have delayed until now the discussion of the important issue of the
{\it gauge problem}. In fact, the 'MiSaTaQuWa' force
(\ref{MiSaTaQuWa}) is computed in the harmonic gauge.  Since the
'force' is not a gauge invariant concept we need to deal with the
problem of gauge transformations \cite{Barack:2001ph}. In particular,
it is difficult to compute the transformation between a singular gauge
at the location of the particle, as the Regge-Wheeler gauge, and a
regular and locally isotropic one as the harmonic gauge.

\section{The solution to the problems of the solution}

The reason we used the Regge-Wheeler gauge was its convenience for
solving General Relativity's field equations, i.e. reducing them to
solve the Zerilli and Regge-Wheeler wave equations.  The natural way
of proceeding for a computation oriented to the evaluation of the
self-force is to solve the ten General Relativity field equations
\begin{equation}\label{GRfieldEq}
\square\bar{h}_{\alpha\beta}+2R^{\mu\ \nu}_{\ \alpha\ \beta}\bar{h}_{\mu\nu}=S_{\alpha\beta},
\end{equation}
where $S_{\alpha\beta}$ are the source terms due to the orbiting particle,
directly in the {\it harmonic gauge}
\begin{equation}\label{harmonic}
g^{\beta\gamma}\bar{h}_{\alpha\beta;\gamma}=0,~{\rm where}~~
\bar{h}_{\alpha\beta}=h_{\alpha\beta}-\frac12 g_{\alpha\beta} h.
\end{equation}
The main inconvenience to overcome are
the potential numerical instabilities of this formulation.

A proper use of the coordinate conditions (\ref{harmonic}) that
eliminates all the first time derivatives leading to instabilities in
equations (\ref{GRfieldEq}) is possible (Barack and Lousto, in
preparation). We can then use the algorithm of Ref.~\cite{Lousto97b}
to integrate the wave equations with the Dirac's delta source terms
appearing in (\ref{GRfieldEq}).

The advantages of this procedure are apparent: No gauge problem, the
computed force will be in the harmonic gauge; so are the
regularization parameters. We have to deal with $C^0$ quantities
(metric coefficients) as opposed to discontinuous waveforms or
divergent metric coefficients, and will only need first order
derivatives of them to compute the self-force. The procedure is in
principle generalizable to Kerr backgrounds (although the actual
implementation will not be a straightforward one). Finally, the
non-radiative modes $\ell=0,1$ can be solved in practically an
analytic way in the harmonic gauge \cite{Detweiler:2003ci}.

\section{more problems seeking for solutions}

As we mentioned earlier the self-force is not a gauge invariant
concept. A way to recover a physical interpretation of the corrected
trajectory of the particle is to use it to compute the corrections to
the energy and momentum radiated at infinity from the binary
system. In order to do that we have to deal with second order
perturbations of the large black hole background. Including this
metric perturbations is needed to consistently take into account all
${\cal O}(m/M)^2$ terms (the other coming from the corrected
trajectory of the particle.)  To second perturbative order, products
of first order perturbations act as sources \cite{Campanelli99}, as a
result, we could have singular products at the location of the
particle (we only probed regularization of the force, not of each
component of the first order metric and its derivatives).

Rosenthal and Ori have recently worked out a scheme to extract the
leading singular piece of the (quadratic) source term using the
retarded field and a choice of the local gauge to regularize the
problem.  This promising approach may solve the last standing
theoretical problem, but we know by now that new, sometimes
unexpected, problems may still arise in our way to provide the
complete solution.

\ack
The author thanks L.Barack for reading the original manuscript.
We also gratefully acknowledge the support of the NASA Center for
Gravitational Wave Astronomy at The University of Texas at Brownsville
(NAG5-13396) and NSF for grants PHY-0140326 and PHY-0354867.

\vskip 10pt

\providecommand{\bysame}{\leavevmode\hbox to3em{\hrulefill}\thinspace}
\providecommand{\MR}{\relax\ifhmode\unskip\space\fi MR }
\providecommand{\MRhref}[2]{%
  \href{http://www.ams.org/mathscinet-getitem?mr=#1}{#2}
}
\providecommand{\href}[2]{#2}

\end{document}